\documentstyle[prl,aps,epsf,multicol]{revtex}

\begin{document}
\draft
\preprint{{\bf ETH-TH/98-??}}

\title{Transport in a One-Dimensional Superfluid: Quantum Nucleation 
of Phase Slips}

\author{H.P.\ B\"uchler $^{a}$, V.B.\ Geshkenbein $^{ab}$, and G.\
  Blatter $^{a}$}

\address{$^{a}$Theoretische Physik, ETH-H\"onggerberg, CH-8093
  Z\"urich, Switzerland\\ $^{b}$Landau Institute for Theoretical
  Physics, 117940 Moscow, Russia}

\date{\today} 
\maketitle
\begin{abstract}

{We present an analytical derivation for the quantum decay rate of the  
superflow through a weak link in a one-dimensional (1D)
Bose-Einstein-condensate. The effective action for the phase
difference across the link reduces to that of a massive particle with
damping subject to a periodic potential.  We find an algebraic
flow--pressure relation, characteristic for quantum nucleation of
phase slips in the link and show how short-wave length fluctuations
renormalizing the interaction between Bosons remove the quantum phase 
transition expected in this class of systems.} 
\end{abstract}

\pacs{PACS numbers: 
  67.40.Hf, 
  67.40.Rp, 
  74.40.+k, 
  67.40.Kh 
}

\begin{multicols}{2}
\narrowtext

Quantum fluids exhibit fascinating phenomena as they push quantum
mechanical effects to the macroscopic level. Classic examples are the
electronic condensate in superconductors and the uncharged bosonic
$^{4}{\rm He}$ and fermionic $^{3}{\rm He}$ superfluids
\cite{tilley90}.  Furthermore, the new laser- and evaporative cooling
techniques have opened up new opportunities in the fabrication and
manipulation of Bose-Einstein condensates (BEC) in various
alkali-metal vapors \cite{anderson95}. A spectacular phenomenon is the 
friction free transport, often studied in terms of the Josephson
effect across a constriction or orifice separating two
reservoirs. Here, we concentrate on a specific geometry, a narrow
channel, and study the decay of the superflow due to quantum
generation of phase slips. Such decay processes have been studied in 
superconducting wires, both experimentally \cite{giordano94} and
theoretically \cite{zaikin97}. In this letter, we investigate the
quantum decay of driven uncharged condensates --- the Galilean
invariance and the Schr\"odinger dynamics in these systems pose new
challenges in the theoretical description and produce drastic changes 
in the physical results as compared to the charged situation.

In a (thick) one-dimensional superconducting wire,
topological 
fluctuations of the phase (phase slips) are weak, resulting in a true
superconducting response \cite{zaikin97} with an algebraic
current-voltage ($I$-$V$) characteristic $V \propto I^{2 g -2}$ and a
dimensionless coupling constant $g\approx 50 r_{0}/\lambda_{L}$ ($r_{0}$  
denotes the radius of the wire and $\lambda_{L}$ the London penetration
length).  Reducing $g$, e.g., via
decreasing the diameter of the wire, the system undergoes a ($T=0$)
quantum phase transition at $g=2$ to a metallic phase \cite{zaikin97}
with an ohmic characteristic. 
In the metallic wire the dynamics of the phase
derives from integration over fermionic modes and is determined
through the coupling to the electromagnetic field. Here, we study an
uncharged BEC which is accurately described by the Gross-Pitaevskii
Lagrangian for the condensate wave function $\psi(x,t)$. 
The Galilean invariance of the homogeneous system prohibits the
nucleation of phase slips and the homogeneous system is always in the
superfluid state. Including a weak link breaks Galilean invariance and
phase slips can nucleate at the defect \cite{freire97}, however, we
find that the link always remains in the superfluid state and the
entire system exhibits no quantum phase transition whatsoever.
\begin{figure}[p]
\vspace{-0.5cm} 
\centerline{\epsfxsize= 7.0cm \epsfbox{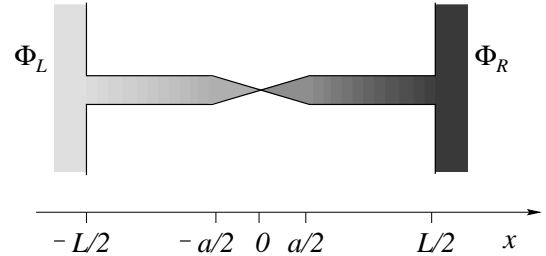}}
\vspace{0.3cm} 
\caption{The 1D channel is connected to two superfluid reservoirs
with a weak link at $x=0$. The superflow $I$ in the channel is driven
by the phase difference $\Delta \Phi=\Phi_{R} -\Phi_{L}$ between the
reservoirs.} 
\label{superchannel}
\end{figure} 
We start with the Gross-Pitaevskii Lagrangian giving the dynamics
of the condensate wave function \cite{dalfovo99}.  With $m$ the mass
of the bosons, $\rho_{0}$ the condensate density, and $U$ the strength 
of the repulsion, the Lagrangian reads
\begin{equation}
  {\mathcal L}_{GP}= i \hbar \: \overline{ \psi} \partial_{\tau} \psi
   - \frac{\hbar^{2}}{2 m } (\partial_{x} \overline{\psi})
   (\partial_{x} \psi) - \frac{U}{2} \left[\overline{\psi} \psi -
   \rho_{0} \right]^{2} \label{grosspitaevskii} \: .  \end{equation}
The system admits states $\psi = \sqrt{\rho_{I}} \exp(i m v x/\hbar)$
carrying a flow $I= \rho_{I} v$ with the renormalized density
$\rho_{I}=\rho_{0} - m v^{2} /2U$ and the velocity $v$, driven 
by the phase difference $\Delta \Phi= m v L/\hbar$ across 
the reservoirs. The suppression of the modulus produces
the critical flow $I_{c} = (2/3)^{3/2} \rho_{0} \sqrt{U\rho_{0}/m}$.  
A crucial feature of the Lagrangian (\ref{grosspitaevskii}) is 
its Galilean invariance: with $\psi(x,t)$ a state, further solutions
are generated by the boost $\exp(i m \left[-vx + i v^{2}
t/2\right]/\hbar) \psi(x-v t,t)$, with $v$ the relative velocity of the   
two coordinate systems.  A flow carrying state is mapped via such a 
boost to a stable state with $I=0$, hence homogeneously flowing 
states are stable.  An inhomogeneity in the tube breaks the symmetry
and quantum nucleation of phase slips bound to the perturbation
appear.  Here, we introduce a weak link (see Fig.~\ref{superchannel})
modeled by a variation in the stiffness,
\begin{equation}
  {\mathcal L}_{w} = (\hbar^{2}c(x)/2 m) \left( \partial_{x}
     \overline{\psi} \right) \left( \partial_{x} \psi \right) \: ,
\end{equation}
where $c(x)$ vanishes for $|x|> a/2$ ($a$ denotes the width of the weak 
link). The action of the channel then reads
\begin{equation}
   {\mathcal S}= {\mathcal S}_{R} + {\mathcal S}_{L} + \int dt
   \int_{-\frac{a}{2}}^{\frac{a}{2}} dx \left[{\mathcal L}_{GP} +
   {\mathcal L}_{w} \right] \: , 
\end{equation}
with $S_{R(L)}$ the contribution of the right (left) lead. Note, 
that the above perturbation is time independent with the weak 
link at rest with respect to the reservoirs. This contrasts with 
Ref.\ \cite{hakim97} where the flow around a moving 
object has been considered.  Applying the mapping $\psi(x,t) 
\rightarrow \exp(- i m v x/\hbar)\psi(x-v t,t)$ and
$\rho_{0} \rightarrow\rho_{I}$ allows for a comparison of our system
with the setup in \cite{hakim97}.

In order to study the quantum behavior of the superfluid we go over to 
the Euclidean action \cite{negele98}
\begin{equation}
  {\mathcal S}_{E} = \int_{-\frac{1}{2} \hbar \beta}^{\frac{1}{2}
    \hbar \beta} d\tau \int dx \left[ \hbar \: \overline{\psi}
    \partial_{\tau} \psi + {\mathcal H}\left[\overline{\psi}, \psi
    \right] \: \right] \: , \label{euclideanaction}
\end{equation}
with $\tau$ the imaginary time, $\beta=1/T$ the inverse temperature, 
 and $\mathcal{H}\left[\overline{\psi}, \psi \right]$ the Hamiltonian 
 density,
\begin{displaymath}
{\mathcal H}\left[\overline{\psi}, \psi \right] =
\frac{\hbar^{2}\left[1-c(x) \right] }{2 m } (\partial_{x}
\overline{\psi}) (\partial_{x} \psi ) + \frac{U}{2} \left[\,
\overline{\psi} \psi - \rho_{0} \right]^{2} .
\end{displaymath}
A saddle point solution extremizing the above action involves
$\psi$ and $\overline{\psi}$ as independent variables; this 
freedom allows for an analytic continuation of the integration 
contour in the path integral.  Performing functional
derivatives produces the equations
\begin{eqnarray}
   - \hbar \: \partial_{\tau } \psi & =  &  -
   (\hbar^{2}/2 m) \partial_{x}^{2} \psi  +  U 
   \left[\, \overline{\psi} \psi - 
     \rho_{0} \right] \psi  \: ,\nonumber \\
   \hbar \: \partial_{\tau } \overline{\psi} & =  &  -
   (\hbar^{2}/2 m) \partial_{x}^{2} 
   \overline{\psi}  + U  \left[\, \overline{\psi} \psi - 
     \rho_{0} \right] \overline{\psi} 
   \: .
   \label{psibarequation}
 \end{eqnarray} 
Note that with $\psi(\tau)$, $\overline{\psi}(\tau)$ satisfying
Eq.~(\ref{psibarequation}) the configuration
$\overline{\psi}\:^{*}(-\tau)$, $\psi^{*}(-\tau)$ is also a solution, 
with $\psi^{*}$ the complex conjugate of $\psi$. Furthermore, the 
periodicity of $\psi$ and $\overline{\psi}$, i.e., $\psi(x,\tau+\hbar 
\beta)=\psi(x,\tau)$, and the boundary condition $\psi^{*}(x,\hbar
\beta /2) = \overline{\psi}(x,\hbar \beta/2)$ imply that
 \begin{equation} \overline{\psi}(\tau) = \psi^{*}(-\tau)
   \label{saddlepointrelation} \end{equation}
for the saddle point solution. 
 
The 1D quantum phase slips, saddle point solutions describing the
quantum tunneling of the $\psi$ field, take the topological form of
vortex-antivortex pairs in the $(x,\tau)$-plane \cite{zaikin97}. Within 
the vortex core region the modulus of the wave function drops to zero,
while in the region far from the vortex center its variation 
is small and the action is determined by the behavior of the
phase. In the uncharged superfluid, Galilean invariance prohibits
formation of a vortex-antivortex pair in the homogeneous region and
thus the vortices are bound to the weak link. In the following we
derive an effective action for the phase difference across the link.

We first integrate out the weak link.  For a small width $a< \xi$ the 
dominant contribution in the action is the kinetic term
\cite{aslamazov68} and the equation for the wave function $\psi$
inside the weak link simplifies to
\begin{equation} 
  \partial_{x} \left\{ \left[1- c(x) \right] \partial_{x} \psi
  \right\} = 0
  \:.  \label{weaklinkequation} 
\end{equation}
The boundary conditions are $\psi(a/2) = \psi_{R}(a/2)$ and $
\psi(-a/2,\tau ) = \psi_{L}(-a/2)$ with $\psi_{R} $ ($\psi_{L}$) the
wave function in the right (left) lead. Inserting the solution 
of (\ref{weaklinkequation}) into the action we obtain the contribution  
\begin{equation}
   \frac{{\mathcal S}_{w}}{\hbar}= \int d\tau \frac{I_{w}}{2 \rho_{0}}
     \left[ \overline{\psi}_{L} \psi_{L} + \overline{\psi}_{R}
     \psi_{R} -\overline{\psi}_{R} \psi_{L}- \overline{\psi}_{L}
     \psi_{R } \right] \:.  \label{weaklinkaction}
\end{equation}
Note, that in the above expression $\psi_{R}$ and $\psi_{L} $ are
taken at $x=a/2$ and $x=-a/2$. The critical flow of the weak link
$I_{w}^{-1} = (m/\hbar \rho_{0}) \int_{-a/2}^{a/2} dx
\left[1-c(x)\right]^{-1}$ is assumed to be small compared to the
critical flow $I_{c} $ of the homogeneous system $I_{w}/I_{c} \ll 1 $. 

Next, we focus on the contribution of the leads. The flow in the wire is 
limited by the small  critical current $I_{w}$ of the link, justifying the
ansatz   
\begin{equation}
  \psi = \sqrt{\rho_{0}} \: \left( 1 + h \right) \: e^{i \phi}
  \label{ansatz} 
\end{equation}
with $h$, $\partial_{x}\phi$, and
$\partial_{\tau} \phi$ small, of order $I_{w}/I_{c}$. For the
following calculations, it is convenient to define the length $\xi=
\hbar / \sqrt{m \rho_{0} U}$, the sound velocity $v_{s}^{2}= \rho_{0}
U/m$, and the time $\tau_{0} = \xi / v_{s}$. Inserting the ansatz
(\ref{ansatz}) into (\ref{psibarequation}), the modulus and phase
decouple to lowest order in $I_{w}/I_{c}$, and using 
(\ref{saddlepointrelation}) the equation for the phase
$\phi$ takes the form
\begin{equation}
   - \tau_{0} \: \partial_{\tau} \phi(x,\!\tau) + (\xi^{2}/2) \:
   \partial_{x}^{2} \phi(x,\!\tau)\! = \!\sin \left[ \phi^{-}(x,\! \tau)\right],  
\label{phiequation}
\end{equation}
with $\phi^{-}(x,\tau)= [ \phi(x,\tau) -\phi(x,-\tau) ]/2$.  The left 
side defines a diffusion equation, while the right side describes a
nonlinear and nonlocal coupling of the wave function. The boundary
conditions demand that the asymmetry $\phi^{-}$ vanishes in the limit
$x\rightarrow \infty$ and $\tau \rightarrow \pm \hbar \beta/2$, see
Eq.~(\ref{saddlepointrelation}).
The solution of the linearized system then is a promising ansatz
for describing the solution far from the vortex core region. Below
we find that the asymmetry is of order $I_{w}/I_{c}$ outside the
core region and the relevant behavior of the phase slip is determined
by the region where the linearization of Eq.~(\ref{phiequation}) is
justified.  Applying a Fourier transformation leads to a local and
linear system of equations for $\phi(\omega)=u(\omega)+iv(\omega)$
\begin{equation}
 - \xi^{2} k^{2} \left( \begin{array}{c} u \\ v \end{array}
   \right) = \mathcal{A} \left( \begin{array}{c} u \\ v \end{array}
   \right), \hspace{6pt} \mathcal{A} = \left[ \begin{array}{c c} 0 &
   - 2 \omega \tau_{0} \\ 2 \omega \tau_{0} & 4 \end{array} \right] ,
   \label{coupeldequation} 
\end{equation}
defining modes with a dispersion relation $- \omega^{2}/v_{s}^{2} = 
k^{2} \left(1 + \xi^{2} k^{2}/4 \right)$. Next, we solve 
(\ref{coupeldequation}) for the phase in the right lead $x>a/2$ 
with the boundary conditions $\phi_{R}(x\rightarrow\infty,\tau)=0$ 
and $\phi_{R}(a/2,\omega)=p(\omega)+iq(\omega)$, where 
$\psi_{R}(x,\tau)=\exp(i\phi_{R}(x,\tau))$,
\begin{eqnarray}
   u(x,\omega) & = & \frac{1}{\lambda_{-} - \lambda_{+}}
     \left\{\frac{}{} \left[ \lambda_{-}\: p + 2 \omega \tau_{0} \: q 
     \right] e^{-s_{+} (2x-a) /2\xi} \right.
   \!\!\!\label{ugeneralsolution} \\ 
   & & 
     \hspace{37pt} \left. \frac{}{}- \left[ \lambda_{+} \: p + 2
     \omega \tau_{0}\: q \right] e^{- s_{-} (2x-a) /2 \xi} \right\} \: 
 ,\hspace{-10pt} 
     \nonumber \\ v(x,\omega) & = & \frac{- \lambda_{+}
       \lambda_{-}}{\lambda_{-} - 
     \lambda_{+}} \left\{ \frac{ 
     \lambda_{-} \: p + 2 \omega \tau_{0} \: q }{2 \omega \tau_{0}
   \lambda_{-}} e^{-s_{+}(2 x-a) /2
     \xi} \right. \label{vgeneralsolution}  
 \\ & & \hspace{42pt}
     \left. - \frac{\lambda_{+}
     \: p + 2 \omega \tau_{0} \: q }
   {2 \omega \tau_{0} \lambda_{+}} e^{- s_{-}
     (2x-a) /2 \xi} 
     \right\} \: ,  \nonumber
\end{eqnarray}
with $\lambda_{\pm}= 2 ( 1 \pm \sqrt{1-\omega^{2} \tau_{0}^{2}})$ and $ 
s_{\pm} = \sqrt{\lambda_{\pm}}$. Inserting the solution back into 
the effective action ${\mathcal S}_{E}$, Eq.~(\ref{euclideanaction}),
we concentrate on the first term (the second term involves the 
Hamiltonian $H= \int dx {\mathcal H}$ which is conserved under 
time evolution and therefore does not contribute). Linearizing
in the small quantities $h$, $\partial_{x}\phi$, and 
$\partial_{\tau}\phi$, the variation $h$ of the modulus drops out 
and the action is determined by the phase field alone,
\begin{equation}
   {\mathcal S}_{E} = - \hbar \rho_{0} \int_{-\infty}^{\infty}
   \frac{d\omega}{\pi} \int_{a/2}^{\infty} dx \: \omega \: u(x,\omega)  
   \: v(x,\omega) \: .  \label{FTaction}
\end{equation}
A straightforward calculation leads to the effective action for the
phase $\phi_{R}(a/2,\tau)$ on the right side of the weak link
 \begin{displaymath} {\mathcal S}_{E} = \frac{\hbar K}{\pi}
   \int_{0}^{\infty} \frac{d\omega}{4 \pi} \frac{\omega}{(1+\omega
   \tau_{0})^{1/2}} \left[ (p-q)^{2} \frac{2+\omega \tau_{0}}{1+\omega 
   \tau_{0}} - 2 q^{2}\right] \end{displaymath}
with $K= \pi \rho_{0} \xi$ the number of particles per coherence
length.  An expansion in $\omega \tau_{0}$ is equivalent to an
expansion in $I_{w}/I_{c}$ and we obtain to lowest order
\begin{equation}
   {\mathcal S}_{E} = \frac{\hbar K}{\pi} \int_{-\infty}^{\infty}
   d\omega \left[ \: |\omega| p^{2} - 2 \omega p q \right] \: .
   \label{rightaction}
\end{equation}
The first term is a Caldeira-Leggett type damping \cite{caldeira83} for 
the symmetric part $p$ of the phase $\phi_{R}(a/2,\tau)$ and describes 
the flow of energy from the weak link to the reservoirs via sound 
wave excitations. Such a term is expected for a dynamics admitting 
sound waves and is also present in the charged superconductor.  
The second term is a coupling between symmetric 
and asymmetric parts and is only present for Schr\"odinger dynamics. 

An analogous calculation with $\phi_{L}(-a/2,\tau)$ adds a similar 
effective action for the phase on the left side of the link. Adding
up terms from the leads and the link we obtain the total
effective action for the phase difference $\varphi(\tau) =
\phi_{R}(a/2,\tau) -\phi_{L}(-a/2,\tau)$ across the weak link
(minimization with respect to the `center of mass' variable 
$\vartheta(\tau)=\phi_{R}(a/2,\tau)+\phi_{L}(-a/2,\tau)$ gives 
$\vartheta=0$). Introducing the notation $\varphi^{\pm} =
\left[\varphi(\tau)\pm\varphi(-\tau)\right]/2$ for the symmetric 
and asymmetric parts we obtain
\begin{eqnarray} 
\frac{{\mathcal S}_{\rm eff}}{\hbar}\!&=&\!
\frac{K}{2\pi}\!\int\!\!d\tau \Biggl[
                \int\!\!d\tau'
    \frac{\left[\varphi^{+}(\tau)\!-\!\varphi^{+}(\tau')\right]^2}
         {4\pi\left(\tau-\tau'\right)^2} 
  -\varphi^{-}(\tau)\partial_{\tau}\varphi^{+}(\tau)
                          \Biggr]  \nonumber \\
   && \hspace{-25pt}
  +\!\int\! d\tau \{I_{w}\left[1-\varphi^{-}(\tau)^{2}/2
  -\cos\varphi^{+}(\tau)\right] - I\varphi^{+}(\tau)\}. 
\label{schjunctioneffaction}
\end{eqnarray}
Integrating out $\varphi^{-}$ in ${\mathcal S}_{{\rm eff}}$  produces 
the mass term $\hbar\int d\tau  \: m_{w} (\partial_{\tau}
\varphi^{+})^{2}/2$ with $m_{w}=K^{2}/4I_{w}\pi^{2}$ via the relation  
$\varphi^{-} = -(K/ 2 \pi I_{w}) \:  
\partial_{\tau} \varphi^{+}$.  
Collecting terms, the effective action for the symmetric part 
$\varphi^+$ of the phase difference across the link is equal to the 
action for a particle with mass $m_{w}$ and
damping $\eta = K/2 \pi$ in a periodic potential and driven by the
force $I$.  The general form of a phase slip is $\varphi^{+}(\tau) =
g^{+}(\tau) + g^{-}(\tau) + \arcsin(I/I_{w})$ where the shift described 
by last term is due to the driving force $I$, while $g^{+}$ ($g^{-}$)
denotes a kink (anti-kink) associated with the presence of a vortex
(anti-vortex) in the link.  Applying the trial function $g^{\pm}= \pm
2 \arctan\left[(2 \tau\pm \overline{\tau})/\tau_{c}\right]$ with
the kink width $\tau_{c} = K/\pi I_{w}$ \cite{korshunov87}, the
action for the phase slip takes the form
%
\begin{equation}
   {\mathcal S}_{{\rm eff}}/\hbar = 2 K \ln \left(
      \overline{\tau}/ \tau_{c} \right) -  2 \pi I 
    \overline{\tau}   - \ln y \label{trialaction} \: , 
\end{equation}
%
where $\ln y$ is found by collecting terms independent of
$\overline{\tau}$. The kinks interact logarithmically with strength 
$2K$, while the driving force $I$ produces a linear repulsion 
within a pair. Using the above trial function $g^{\pm}$, we can 
justify the linearization in (\ref{phiequation}): the region
of large asymmetry $\phi^{-}(x,\tau)$ is limited by $\xi$ along $x$
\cite{bem1} and by the width $\tau_{c}$ of the kink along the
$\tau$ direction. The contribution from the core then is
independent of $\overline{\tau}$ and only enters through the fugacity  
$y$.

{\em Thermodynamics} ($I=0$): In the following we consider a gas of
quantum phase slips. The associated action describes charged
particles with logarithmic interaction in 1D and a chemical potential
$\ln y$. The one-dimensionality of the gas follows from the confinement 
of the vortices to the weak link. This contrasts with the 
superconducting wire where phase slips appear throughout the 
homogeneous channel and the action of the quantum phase slip gas 
is mapped to that of 2D charged particles with logarithmic 
interaction of strength $2 g$ \cite{zaikin97}. 

In the latter system, a quantum phase transition (at $g=2$) has 
been found \cite{zaikin97} as quantum fluctuations drive the system 
ohmic at small values of $g \leq 2$ (including a weak link and ignoring 
phase slips in the wire shifts this transition to $g=1$ \cite{glazman}). We
then may expect a similar  
transition to occur in the superfluid at $K=1$.  Indeed, a careful 
analysis shows that a massive particle with damping and subject 
to a periodic potential undergoes a quantum phase transition for 
a critical damping $K=1$ \cite{schmid83}. However, in the above
derivation of the effective action (\ref{schjunctioneffaction}) 
we have neglected high frequency fluctuations of the order parameter 
which affect the long wave length behavior of the homogeneous 
superfluid via a renormalization of the sound velocity $v_{s} 
=\sqrt{f\left(\gamma \right)} \rho_{0} \hbar/m$, where $\gamma
= m U/\hbar^{2} \rho_{0}$. The universal function $f(\gamma)$ 
is obtained from the exact (Bethe Ansatz) solution of interacting 
1D Bosons \cite{lieb63.1}, 
%
\begin{equation}
   f(\gamma) = \left\{ 
     \begin{array}{cl}
       \gamma - \gamma^{3/2}/2 \pi \: , & \gamma \ll 1 \: , \\  
       \pi^{2} \left(1 - 8/\gamma  \right) \: , & \gamma \rightarrow
       \infty \: .
     \end{array} \right. 
\end{equation}
As the original relation $\pi\rho_{0} \hbar/m =K v_{s}$ between the
sound velocity and the dimensionless parameter $K$ is not 
renormalized, a consequence of the Galilean invariance \cite{haldane81},
the renormalization of $K$ can be found from that of the sound velocity.  
For $\gamma \ll 1$, the result $K=\pi \rho_{0} \xi\gg 1$ is recovered,
while in the limit $\gamma\rightarrow \infty$ we obtain hard-core
bosons with $K=1$ \cite{bem2}. For intermediate values of $\gamma$,
$K$ varies smoothly between these two limits. Therefore, no quantum
phase transition is present in the uncharged superfluid.

\begin{figure}[htb]
\vspace{-0.3cm} 
\centerline{\epsfxsize= 7cm \epsfbox{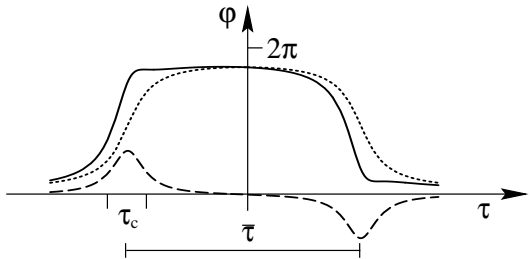}}
\vspace{0.1cm}
\caption{ Phase slip solution: the dotted line is
the symmetric part $\varphi^{+}$, while the dashed line represents
the asymmetric part $\varphi^{-}$. The width of the asymmetry is
determined by $\tau_{c}$ and the distance between the kink --
anti-kink pair is $\overline{\tau}$.}  
\label{phaseslip}
\end{figure}

{\em Superflow} ($I\neq 0$): The shape of the phase slip is determined
by the driving current $I$: minimizing the ($T=0$) action 
(\ref{trialaction}) with respect to $\overline{\tau}$ we find 
the kink--anti-kink separation $\overline{\tau}=K/\pi I$ (see 
Ref.\ \cite{korshunov87} for a detailed discussion addressing
the motion of a particle with damping in a periodic potential).
The kink width $\tau_{c}$ determines the asymmetry of the phase 
slip via the relation $\varphi^-\propto-\partial_{\tau}\varphi^+$, 
see Fig. \ref{phaseslip}. This form of the phase slip agrees 
well with numerical simulations \cite{freire97}.
The quantum decay rate of the superflow is determined by $\Gamma = 
B \exp(- S_{{\rm ps}}/\hbar)$ \cite{callan77} with $S_{\rm{ps}}$ 
the action of the phase slip and $B$ arising from fluctuations around 
the saddle point. At $T=0$, the action is dominated by the
logarithmic interaction between the kink and the anti-kink leading to
a factor $I^{2K}$. Gaussian fluctuations in the
distance between the kink and anti-kink with frequency $\propto I$ lead to 
a preexponential factor $B \propto 1/I$ and therefore $\Gamma
\propto I^{2K-1}$ (in the superconducting wire the relative positions
of the kink -- anti-kink pair fluctuate both along $x$ and $\tau$ and
therefore $B \propto 1/I^{2}$, leading to $\Gamma \propto I^{2 g-2}$). 
Note, that the prefactor produces a sizeable correction of the flow rate
as compared to the saddle point result \cite{freire97}.
At finite temperature we have to determine the transition rates 
$\Gamma^{\pm}$ both to lower and higher energy states. These transitions 
then produce a time dependent phase difference between the reservoirs 
which in turn determines the drop $\Delta \mu$ in the chemical potential 
across the channel,
\begin{equation}
\frac{\Delta \mu}{\hbar} = 2 \pi \left[\Gamma^{+} -\Gamma^{-}\right]
\propto 
\left\{\begin{array}{cl}
       {\displaystyle I_{w}\left(\frac{I}{I_{w}}\right)^{2K-1}\!\!,} &
       {\displaystyle I\gg\frac{T}{\hbar}\: ,\!\!\!}   \\ 
       {\displaystyle I\left(\frac{T}{\hbar I_{w}}\right)^{2K-2}\!\!,}&
       {\displaystyle I\ll\frac{T}{\hbar}\: . \!\!\!}  
       \end{array} \right. 
\label{algebraicflowpressure}
\end{equation}
With the pressure $p \propto \Delta \mu$ we arrive at an algebraic
flow-pressure characteristic for $T=0$, while at $T \neq 0$ the
channel always exhibits a linear behavior for a small superflow $I$.

In the end, we have found a remarkable similarity in the response 
of charged and uncharged condensates: the quantum decay of the 
superflow due to phase slips leads to algebraic current-voltage 
and flow-pressure relations, respectively. On the other hand, we 
find distinct differences as well: first, the quantum nucleation 
of phase slips in the present uncharged situation is bound 
to a defect, a consequence of Galilean invariance, while the same 
phenomenon in the charged case can take place throughout the wire.
Second, the presence of high energy modes renormalizing the sound 
velocity removes the quantum phase transition for the uncharged 
superfluid, preserving the superflow.

We wish to thank K.\ Le Hur and R.\ Heeb for valuable discussions.

\end{multicols}


\begin{thebibliography}{10}
\bibitem{bem1} Eq.\ (\ref{vgeneralsolution}) shows that the asymmetric
part satisfies $v \sim \tau_{0} \omega u$ for $|x|> \xi$ instead of $v
\sim \omega u/I_{w}$ and therefore is suppressed by $I_{w}/I_{c}$.
\bibitem{bem2} This limit is in agreement with the known fact that
hard-core bosons can be mapped to free fermions, while bosonization of
free fermions leads to a Luttinger liquid with $K=1$.
\end{thebibliography}
\end{document}